\documentclass[aip, apl, reprint, floatfix]{revtex4-1} %reprint; preprint
\usepackage{graphicx}% Include figure files
\usepackage{dcolumn}% Align table columns on decimal point
\usepackage{bm}% bold math
\usepackage{mathptmx}
\usepackage{etoolbox}
\usepackage{amsmath}   % For advanced math formatting
\usepackage{amssymb}   % For mathematical symbols
\usepackage[mathlines]{lineno}% Enable numbering of text and display math
\usepackage[utf8]{inputenc}
\usepackage{hyperref}
\hypersetup{colorlinks=true, linkcolor=blue, citecolor=red, urlcolor=cyan}

\usepackage{cleveref}
\usepackage{soul}
\usepackage{hhline} %double line for the table
\usepackage[normalem]{ulem}
\usepackage{siunitx}
\usepackage{placeins} 
\renewcommand{\thefootnote}{\fnsymbol{footnote}}
\newcommand{\ConstRn}{6.1} %kOhm
\newcommand{\ConstIc}{17.1} %nA
\newcommand{\ConstIcRn}{104} %uV
\newcommand{\Constdelta}{0.38} %meV
\newcommand{\beginsupplement}{
    \setcounter{table}{0}
    \renewcommand{\thetable}{S\arabic{table}}
    \setcounter{figure}{0}
    \renewcommand{\thefigure}{S\arabic{figure}}
    \setcounter{equation}{0}
    \renewcommand{\theequation}{S\arabic{equation}}
}

\begin{document}
\preprint{}

\title{Scaffold-Assisted Window Junctions for Superconducting Qubit Fabrication}

%List of the author
%https://docs.google.com/spreadsheets/d/1R8RUL-PpCQrpfyMimjnLN-aaV3XiWA_sxg76ZtMJmmI/edit?usp=sharing

\author{Chung-Ting Ke*}
\affiliation{Institute of Physics, Academia Sinica, Taiwan}
\affiliation{Center for Critical Issues, Academia Sinica, Taiwan}
\author{Jun-Yi Tsai*}
\affiliation{Institute of Physics, Academia Sinica, Taiwan}
\author{Yen-Chun Chen*}
\affiliation{Center for Critical Issues, Academia Sinica, Taiwan}
\author{Zhen-Wei, Xu}
\affiliation{Center for Critical Issues, Academia Sinica, Taiwan}
\author{Elam Blackwell}
\affiliation{Department of Physics, University of Wisconsin-Madison, WI, USA}
\author{Matthew A. Snyder}
\affiliation{Department of Physics, University of Wisconsin-Madison, WI, USA}
\author{Spencer Weeden}
\affiliation{Department of Physics, University of Wisconsin-Madison, WI, USA}
\author{Peng-Sheng Chen}
\affiliation{Industrial Technology Research Institute, Taiwan}
\author{Chih-Ming Lai}
\affiliation{Industrial Technology Research Institute, Taiwan}
\author{Shyh-Shyuan Sheu}
\affiliation{Industrial Technology Research Institute, Taiwan}
\author{Zihao Yang}
\affiliation{Applied Materials, USA}
\author{Cen-Shawn Wu}
\affiliation{Department of Physics, National Changhua University of Education, Taiwan}
\affiliation{Center for Critical Issues, Academia Sinica, Taiwan}
\author{Alan Ho}
\affiliation{Qolab, WI, USA}
\author{R. McDermott}
\email{robert@qolab.ai}
\affiliation{Qolab, WI, USA}
\affiliation{Department of Physics, University of Wisconsin-Madison, WI, USA}
\author{John Martinis}
\email{john@qolab.ai}
\affiliation{Qolab, WI, USA}
\affiliation{Department of Physics, University of California Santa Barbara, USA}
\author{Chii-Dong Chen}
\email{chiidong@gate.sinica.edu.tw}
\affiliation{Institute of Physics, Academia Sinica, Taiwan}
\affiliation{Center for Critical Issues, Academia Sinica, Taiwan}

\date{\today}

\begin{abstract}
The superconducting qubit is one of the promising directions in realizing fault-tolerant quantum computing (FTQC), which requires many high-quality qubits. To achieve this, it is desirable to leverage modern semiconductor industry technology to ensure quality, uniformity, and reproducibility. However, conventional Josephson junction fabrication relies mainly on resist-assistant double-angle evaporation, posing integration challenges. Here, we demonstrate a lift-off-free qubit fabrication that integrates seamlessly with existing industrial technologies. This method employs a silicon oxide (SiO$_2$) scaffold to define an etched window with a well-controlled size to form a Josephson junction. The SiO$_2$, which has a large dielectric loss, is etched away in the final step using vapor HF leaving little residue. This Window junction (WJ) process mitigates the degradation of qubit quality during fabrication and allows clean removal of the scaffold. The WJ process is validated by inspection and Josephson junction measurement. The scaffold removal process is verified by measuring the quality factor of the resonators. Furthermore, compared to scaffolds fabricated by plasma-enhanced chemical vapor deposition (PECVD), qubits made by WJ through physical vapor deposition (PVD) achieve relaxation time up to $\SI{57}{\us}$. Our results pave the way for a lift-off-free qubit fabrication process, designed to be compatible with modern foundry tools and capable of minimizing damage to the substrate and material surfaces.

\end{abstract}


\maketitle %\maketitle must follow title, authors, abstract and \pacs

\section{Introduction}
Realizing fault-tolerant quantum computation (FTQC) requires a large number of high-quality qubits that allow efficient quantum error correction algorithms\cite{mohseni_how_2024, bravyi_future_2022, mohseni_how_2024, shor_fault-tolerant_1996, muthusubramanian_wafer-scale_2024}. Among various qubit platforms, superconducting qubits are particularly promising because of their scalability and relatively simple material requirements\cite{ezratty_perspective_2023, arute_quantum_2019}. Recent demonstrations of break-even threshold with surface codes underscore the critical role of qubit quality in advancing the fidelity of the quantum processing unit (QPU)\cite{google_quantum_ai_and_collaborators_quantum_2024}. The contributions to qubit decoherence include two-level systems (TLS) defects \cite{mcdermott_materials_2009, bejanin_interacting_2021, li-chung_ku_decoherence_2005, graaf_two-level_2020, martinis_decoherence_2005}, quasiparticle poisoning \cite{catelani_quasiparticle_2011, liu_quasiparticle_2024}, and charge noise \cite{schreier_suppressing_2008, wilen_correlated_2021}. While quasiparticle poisoning can be mitigated by improved shielding and gap engineering \cite{catelani_quasiparticle_2011, liu_quasiparticle_2024} and qubit sensitivity to charge noise can be suppressed by appropriate choice of circuit parameters \cite{schreier_suppressing_2008}, qubit energy relaxation induced by TLS remains a roadblock to realizing FTQC. %Therefore, eliminating unwanted TLSs is essential to improve the superconducting qubits' quality and reduce the overhead for the error correction algorithm. 
The key component in the construction of superconducting qubits is the Josephson junction, which provides the nonlinearity needed to implement a qubit but is highly susceptible to TLS-induced degradation \cite{osman_simplified_2021, wu_overlap_2017,anferov_improved_2024, damme_advanced_2024, wan_fabrication_2021, muthusubramanian_wafer-scale_2024, bilmes_-situ_2021}. Given the growing demand for both the quality and quantity of qubits, it is critical to develop a Josephson junction fabrication process that minimizes TLS associated with amorphous interfaces and polymer residues and that is compatible with  industrial-standard wafer-scale technology\cite{anferov_improved_2024, damme_advanced_2024}.

The canonical method for fabricating Josephson junctions involves a double-angle evaporation process, where a polymer resist mask is used and subsequently lifted off after metal deposition \cite{muthusubramanian_wafer-scale_2024,  muthusubramanian_wafer-scale_2024, bilmes_-situ_2021}. However, neither double-angle evaporation nor the lift-off process is compatible with industrial microfabrication due to the requirements of cleanness and high yield. More advanced fabrication methods including trilayer \cite{wan_fabrication_2021} and overlap junction processes \cite{verjauw_path_2022, damme_advanced_2024} are among the attempts to replace the double-angle evaporation technique. A recent demonstration of qubits incorporating overlap Josephson junctions utilized state-of-the-art \SI{300}{mm} semiconductor tooling\cite{damme_argon-milling-induced_2023, damme_advanced_2024}. Here, Josephson junction formation involves etch of the base electrode followed by a separate vacuum step including \textit{in situ} clean of the base metal and oxidation prior to counterelectrode deposition. While the uniformity of the junction resistances was excellent, the global argon plasma cleaning step can damage the exposed metal-air and substrate-air interfaces, leading to the formation of TLSs. The inevitable large area of oxidation creates unwanted oxide layers between the superconducting metals. Therefore, developing an alternative process flow to prevent surface damage and unwanted oxidation layers on the superconducting circuit may further improve qubit quality. 

% this paragraph was originally highlighted. Remove due to the compiling error.
Here, we demonstrate a window junction (WJ) process using a SiO$_2$ scaffold that can be removed in the final stage. The junction is defined by etching a sub-micron window via into SiO$_2$ scaffold, which is a well-developed process with good reproducibility. The WJ process has several advantages over previous approaches. First, when the junction via is defined using state-of-the-art DUV lithography, the WJ process provides excellent control over the junction area. In addition, cleaning of the native oxide of the base electrode is restricted to the area of the junction window, whereas in an overlap junction process the cleaning step can amorphize the substrate in the vicinity of the junction, leading to enhanced loss. For the devices studied here, the area of the chip exposed to the junction plasma clean is a small fraction of order 5$\times 10^{-6}$ of the total $\SI{1}{cm} \times \SI{1}{cm}$ chip area, greatly reducing the potential for damage from the cleaning step. In order to avoid dielectric loss from the SiO$_2$ scaffold, it is necessary to remove the scaffold layer as a final step in device fabrication. Here, we use a vapor hydrogen fluoride (vHF) etch process to remove the sacrificial SiO$_2$ layer, leaving a suspended airbridge to connect the junction counterelectrode to ground \cite{dunsworth_method_2018, ritala_studies_2010}. For optimized SiO$_2$ deposition and etch processes, we find no difference in internal quality factor ($Q_i$)\cite{mcrae_materials_2020}  for resonators fabricated with and without the sacrificial oxide. 

In the WJ process, it is necessary to clean the metal surface in the via prior to oxidation. Although there is concern that cleaning might damage the metal surface, e.g., particularly involving Ar ion milling, previous experiments on liftoff junctions have not shown significant degradation \cite{damme_argon-milling-induced_2023, damme_advanced_2024}. Our proposed method mitigates the risks associated with wafer-scale Ar bombarding that may cause more TLSs on the surface. In addition, it is possible for SiO$_2$ resputtering inside the window or leave SiO$_2$ residue during the Ar milling step. We overcome the concerns by showing good Josephson junction properties with a low sub-gap leakage current. We further show a good reproducibility of the junction resistance, which has the potential to be improved by tuning the fabrication parameters. Lastly, the qubit relaxation time $T_1$ is up to $\SI{57}{\us}$, potentially limited by surface loss arising from the capacitance between the top and bottom Al electrodes. The comparison of different SiO$_2$ reveals the effect of the capacitance to $T_1$. One may expect to improve $T_1$ time by further reducing the parasitic capacitance of the counter electrode regarding the ground. Nevertheless, our results provide a pathway to realize a qubit fabrication technique using a SiO$_2$ scaffold-assisted window to form the Josephson junction, which is compatible with current foundry fabrication sets with minimum damage to the superconductor and silicon wafer surface. Therefore, the WJ architecture offers a promising approach in quantum process science to produce high-quality and scalable superconducting qubits for future quantum computers.

\section{Methods}
\subsection{Fabrication flow}
The fabrication process is illustrated in Figure$\,$1 and is described in the following steps. First, a high-resistance ($\SI{20}{\kilo\Omega\cdot\cm}$) 8-inch silicon substrate is processed with standard piranha solution (sulfuric acid: $\SI{30}{\percent}$ hydrogen peroxide$\,$=$\,$3:1) followed by diluted HF ($\SI{2}{\percent}$) in spin rinsing to remove native SiO$_x$ before metal deposition. The wafer is transferred to the deposition system within five minutes to reduce the regrowth of oxide. A $\SI{100}{\nm}$ aluminum (Al) is deposited in an ultra-high vacuum (UHV) sputtering system ($<\SI{10e-7}{mTorr}$) for the base layer. The base Al is then patterned using standard deep ultraviolet (DUV) photolithography as shown in Fig.$\,$1 (a). The wafer is coated with DUV photoresist and processed with standard lithography techniques. After the development, we conducted a $\SI{2.38}{\percent}$ TMAH wet etch process in a spin etcher to construct the base structures, including the resonators, coupling capacitors, and other microstructures required for the qubit chip. The etch process is terminated by a deionized water rinse and then dried in the same spin etcher system. 

\begin{figure*}[htbp]
    \centering
    \includegraphics[width=\textwidth]{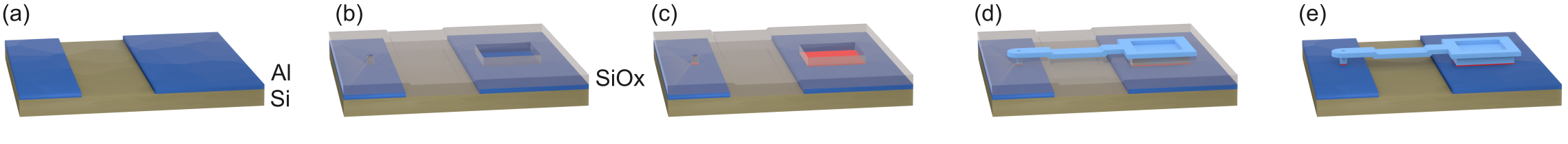}%
    \caption{Illustration of the fabrication process on a high-resistance silicon substrate. (a) The base electrode of $\SI{100}{\nm}$ Al is first deposited, patterned and etched. (b) Next, a $\SI{200}{\nm}$ SiO$_2$ scaffold is deposited on the Al. (c) The junction and contact window are then patterned using DUV lithography, followed by a dry etch to remove the exposed SiO$_2$. Moving the wafer back to the deposition system, a gentle Ar ion cleaning refreshes the surface, followed by an \textit{in-situ} oxidation process and deposition of Al. (d) The top electrode is then patterned. (e) After dicing, the SiO$_2$ scaffold is removed by a vapor HF process. }%
\end{figure*}

Next, the $\SI{200}{\nm}$ SiO$_2$ scaffold layer is deposited on the base electrode, as shown in Fig.$\,$1(b). This scaffold layer deposition is either using the Plasma-enhanced chemical vapor deposition (PECVD) method at $\SI{250}{\degreeCelsius}$ or the Physical vapor deposition (PVD) method by sputtering below $\SI{150}{\degreeCelsius}$. For both types of SiO$_2$, it not only forms the scaffold for junction fabrication but also protects the surface of Al and substrate during fabrication. Another DUV lithography step defines the junction via and a $\SI{10}{}$ by $\SI{10}{\um}$ contact window on the base Al, followed by a fluorine-based dry etching of the SiO$_2$, as depicted by Fig.$\,$1(c).

For junction fabrication, the native AlO$_x$ is removed in both the junction via and the contact window using an argon-based inductively coupled plasma (ICP) etch step ($\SI{450}{W}$ main power and $\SI{300}{W}$ bias power for $\SI{40}{\second}$ under Ar pressure $\SI{5}{mTorr}$). After removing the native AlO$_x$, a high-purity oxygen gas (5N) is introduced to form a controlled AlO$_x$ barrier with a dynamical oxidation process for a time based on the designed junction resistance\cite{muthusubramanian_wafer-scale_2024,  muthusubramanian_wafer-scale_2024, bilmes_-situ_2021}. Note that the area ratio between the contact window and the junction via exceeds 250. Hence, the inductance of the parasitic junction at the contact window is negligible. A $\SI{100}{nm}$ Al layer is deposited \textit{in-situ} immediately after the oxidation process, serving as the counter electrode layer. The Al bridge is again defined by DUV lithography and the TMAH wet etch process, as shown in Fig.$\,$1(d). In this work, the Al top electrode process is conducted in both 4-inch and 8-inch deposition systems showing no significant difference. However, as discussed below, the uniformity of junction oxidation is superior in the 4-inch system. Finally, the SiO$_2$ scaffold is removed by the vHF process prior to bonding and low-temperature characterization. 

\subsection{Window Junction characterization}
We implement several checkpoints to monitor the quality of our via structure and ensure control of our fabrication. These include: (a) the cleanness of the via after the etching step, (b) step coverage of the top Al electrode on the via, (c) variation of the via sizes, and (d) junction resistance uniformity with and without the oxidation process. These checkpoints collectively provide a comprehensive verification of the quality of the WJ fabrication process.

Figure 2(a) shows a typical tilted-angle scanning electron microscope (SEM) image of a junction window.  After removing the photoresist, a well-defined SiO$_2$ via is observed with a clean and damage-free Al base electrode at its bottom, even revealing clean Al grains. We verified good etching for four different via sizes ($\SI{275}{}$, $\SI{300}{}$, $\SI{325}{}$, and $\SI{350}{n\meter}$) with 20 devices for each size. The via diameter, which corresponds to the junction diameter, has variation less than $\SI{2}{\percent}$ (see SI section \ref{sec:ViaFabricationUniformity}).  From the ECDF (integrated histogram) curves, when the via size is reduced to $\SI{275}{n\meter}$ a step feature in the diameters are observed, which becomes sharper at $\SI{350}{n\meter}$. Therefore, the junction size is designed for $\SI{300}{n\meter}$ to $\SI{350}{n\meter}$ in the following qubit fabrication.

To ensure the cleaning process can properly remove the native oxides and does not produce etching residues, we process the junction device for the metallic contact test. The averaged resistance for $\SI{325}{n\meter}$ test junctions is $\SI{5.2}{\Omega}$ with a standard deviation of $\SI{0.39}{\Omega}$, which agrees with the calculated value based on the sheet resistance of Al (SI section \ref{sec:Sup_AlResistance} for other junction dimensions).  
Both etching quality and dimension control support the feasibility of the WJ for future scaling purposes.

With a clean via process, we study the junction oxidation next.  We first studied the oxidized junction structure using transmission electron microscopy (TEM). Figure$\,$2(b) shows the TEM image with a cross-section view of an oxidized WJ before removing the SiO$_2$ scaffold. One important observation is that the top Al electrode has sufficient coverage to fill the via without formation of voids between the two metal layers (see SI \ref{sec:WJTEM-EDX} for EDX analysis). Figure$\,$2(c) further focuses on AlO$_x$, a well-defined oxidized barrier of around \SI{1}{n\meter} is observed between two Al layers in the via. 

Our study uses two different systems to fabricate the top Al electrodes. One is an 8-inch system; therefore, the qubit devices are fully made with an 8-inch process and then diced before vHF. In the 8-inch system, the oxygen gas inlet is located next to the chamber pump-out valve. As it is not possible to throttle the vacuum pump, we can only access low oxidation pressure below $\sim$ $\SI{4.3}{mTorr}$. Moreover, oxygen pressure is not uniform in this system, leading to a large spread of oxygen exposure and junction specific resistance across the wafer. Straightforward modifications to the junction growth chamber will correct this limitation in the future. Another is a 4-inch system, where we found a better oxidation condition because of a proper system design. For the 4-inch case, the 8-inch wafer is diced before the top-electrode process. 

Figure$\,$2(d) shows the empirical cumulative distribution function (ECDF) of the normalized junction resistance of 5 test chips with 14 junctions per chip for \SI{300}{\nm}, \SI{325}{\nm}, and \SI{350}{\nm} via, for the 4-inch process case. A sharp slope in the ECDF plot indicates a small resistance spread around \SI{5.9}{\percent} for the \SI{350}{\nm} via. To further verify the quality of the oxidized junction, we measured the I-V characteristics of the junction in a dilution refrigerator showing a typical I-V curve as presented in Fig.$\,$3(e) indicating the critical current, $I_c=\SI{\ConstIc}{n\ampere}$, and superconducting gap $2\Delta =\SI{\Constdelta}{m\volt}$, and the resistance $R_n= \SI{\ConstRn}{\ohm}$. Based on the Ambegaokare-Baratoff relation, the I$_c$R$_n$ product is $\SI{\ConstIcRn}{\mu\volt}$, consistent with the previous study of the Josephson junction based on aluminum\cite{ambegaokar_tunneling_1963}. Moreover, no excessive leakage current is observed in the subgap, indicating that a high-quality AlO$_x$ barrier can be achieved by via junction using a gentle ion-cleaned Al surface.

\begin{figure}[htbp]
\includegraphics{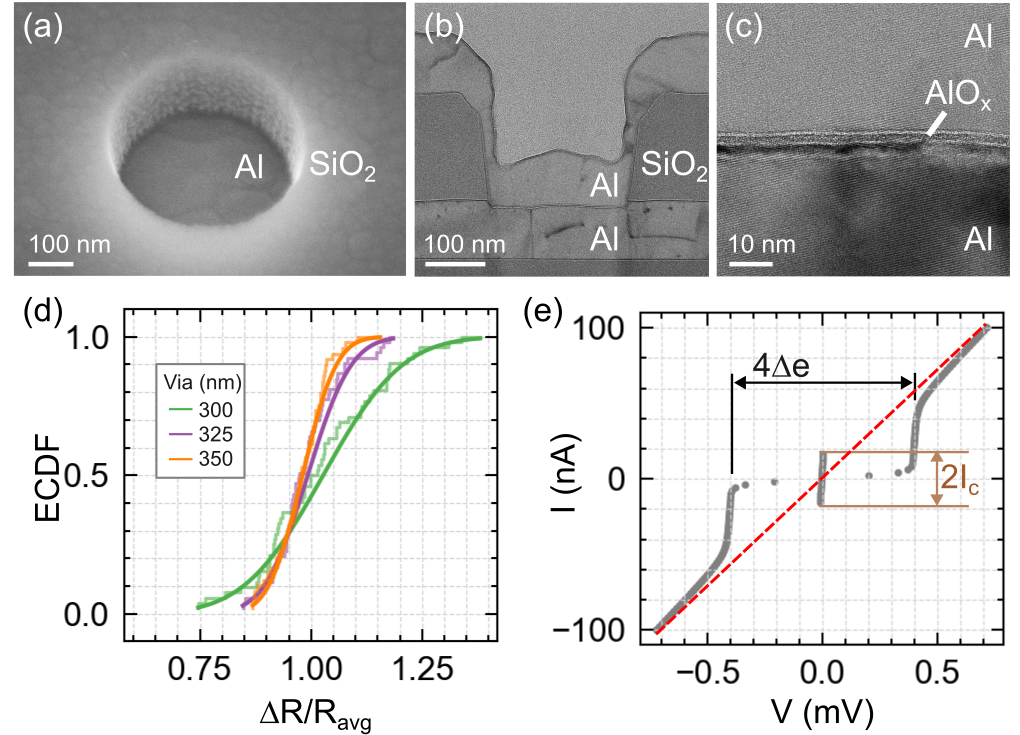}%
\caption{Characterization of the window junction process. (a) A tilted SEM image of the junction via shows a clean Al surface in the window. The Al grains are undamaged after the etching step. (b) Transmission electron microscopy (TEM) image of the window before removing the SiO$_2$ scaffold. (c) TEM image of the junction interface with two aluminum layers and the aluminum oxide tunnel barrier. (d) The room-temperature junction resistance, measured with 4 probes, is plotted as an empirical cumulative distribution function (ECDF). The average resistance(variation) of \SI{300}{}, \SI{325}{}, and $\SI{350}{n\meter}$ junction is \SI{15.3}{k\Omega}\,(\num{15}\,\%), \SI{14.7}{k\Omega}\,(\num{7.7}\,\%), and \SI{13.8}{k\Omega}\,(\num{5.9}\,\%), respectively. Each stepped line is from 14 test junctions on the same chip.  (e) The current-voltage relation of the single test junction at \SI{100}{m\kelvin} and the energy gap $2\Delta =\SI{\Constdelta}{\mV}$ and $I_c=\SI{\ConstIc}{n\ampere}$. As $V>\SI{0.5}{m\volt}$, a linear fit (red dashed line) gives $R_n= \SI{\ConstRn}{\ohm}$.} 
\end{figure}

\section{Results and Discussion}
\subsection{Device quality verification}
One of the essential steps in WJ fabrication is the complete removal of the SiO$_2$ scaffold. Here, we measure the residue of the SiO$_2$ after vHF etch by checking resonator quality factors $Q_i$ before SiO$_2$ deposition and after its removal \cite{mcrae_materials_2020, dunsworth_method_2018}. We fabricated six $\lambda/4$ coplanar resonators with central line widths of $\SI{100}{}$, $\SI{50}{}$, $\SI{20}{}$, $\SI{10}{}$, $\SI{5}{}$, and $\SI{2}{\um}$, maintaining a fixed gap-to-center line width ratio of 0.6 on a single chip. Scaling of $Q_i$ with width and its power dependence is used to confirm surface loss from TLS.  We deposited $\SI{200}{\nm}$ SiO$_2$ using PVD on top of the resonator structures, which is subsequently removed using a precisely calibrated vHF process. 

In Figure$\,$3(a), we show dependence of the resonator quality factor $Q_i$ versus the excitation photon number, along with different widths. The process without and with SiO$_2$ coverage process is labeled with dot and cross markers, respectively, and show little difference. The power and width dependence is typical for high-quality Al \cite{wenner_surface_2011}. Notice that $Q_i$ reaches 1M with a $\SI{20}{\um}$ central-line resonator at the single photon region. The high $Q_i$ shows a low TLS density even when processed with SiO$_2$ deposition and vHF removal, therefore providing strong evidence that removing SiO$_2$ using vHF can realize a high-quality window junction qubit. We can further understand this result based on previous studies in TLS theory that the loss tangent is substantial for the substrate-to-metal interface, which remains intact during this step\cite{wenner_surface_2011}. For the \SI{50}{} and \SI{100}{\um} samples, Q$_i$ fluctuates more due to trapped magnetic field, so the data is not shown. Using SEM, we observed no damage on the Al film after removing the SiO$_2$. The SEM images reveal Al grains that typically observed in the pristine Al film. Atomic force microscopy(AFM) shows an RMS roughness around \SI{1.5}{\nm}, which is comparable to that of the pristine Al film after deposition with roughness around \SI{1.2}{\nm}(The detailed discussion see SI \ref{sec: AlRoughness}). However, we note that the roughness may be influenced by residual SiO$_2$ which may not affect the Q$_i$ due to the small loss tangent of the metal-air interface. 

We check qubit devices manufactured using the WJ process by examining the surface morphology of the junction structure using a SEM. In Fig.$\,$4(a), a tilted angle SEM image shows a well-defined Al pillar from the window junction. A clean surface is observed after the removal of the SiO$_2$ scaffold. In addition, the Al counter electrode is designed and fabricated so that Al remains intact after removing the SiO$_2$ scaffold [Fig.$\,$4(b)]. The top electrode is suspended well without collapsing, where the gap between the electrode and the silicon substrate is around $\SI{200}{\nm}$ corresponding to the thickness of the SiO$_2$ scaffold.
We also compare the SEM image of the SiO$_2$ scaffold using PECVD or PVD, as shown in Fig.$\,$4 (c) and (d), respectively. This clearly indicates that removal of PECVD SiO$_2$ is a problem compared to the PVD process.

\begin{figure}[htbp]
\includegraphics{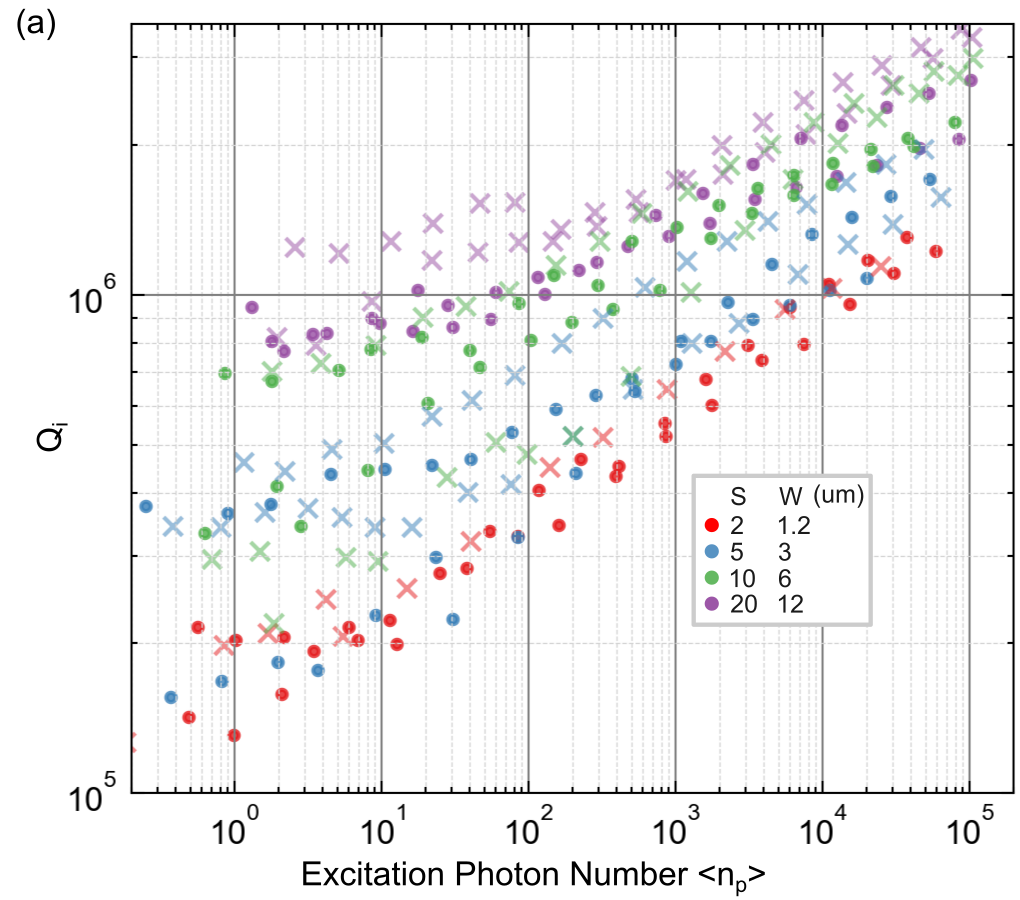}%
\caption{Power dependence of the internal quality factor $Q_i$ of coplanar resonators with widths of the central line S and gaps W; (S, W) = (2, 1.2), (5, 3), (10, 6), (20, 12)$\SI{}{\,\um}$, colored in red, yellow, green, blue, respectively. The dot marker shows data for a typical Al process, as shown in Fig.$\,$1, whereas the cross shows after SiO$_2$ deposition and etch.  This data indicates little residue after vHF.}
\end{figure}

\begin{figure}[htbp]
\includegraphics{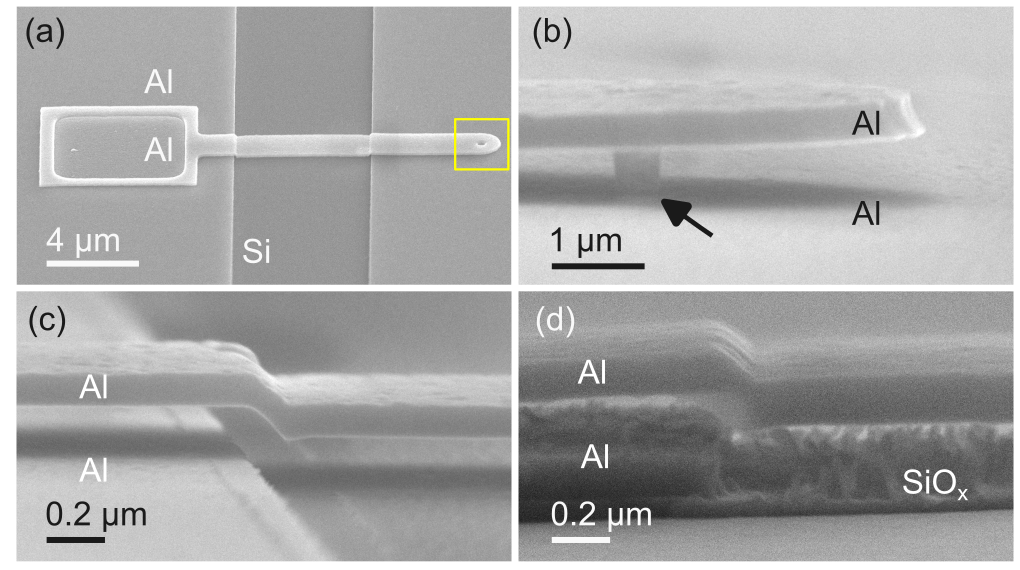}
\caption{(a) Tilted-angle scanning electron microscope (SEM) image of the window junction (WJ).  (b) Magnified view of the junction structure, corresponding to the yellow square in (a); the arrow indicates the junction location. (c) and (d) SEM images of the edge-suspended top electrode fabricated using PVD and PECVD processes, respectively. Residue from the PECVD process remains after the vHF treatment.}
\end{figure}

\subsection{Qubit Characterization}
We characterize the qubit devices in a dilution refrigerator with a base temperature of $\SI{10}{m\kelvin}$, with its measurement setup detailed in SI \ref{sec: MicrowaveMeasurementSetup}. In Table \ref{tab:QubitProperties} we compare the results for SiO$_2$ using PECVD (Q0) and PVD (Q1, Q2). The qubit devices using the PECVD is found to have a short energy decay time $T_1 = 2\,\mu$s, consistent with its large residue.  As shown in Fig.$\,$5(a), Q1 has much longer coherence $T_1 = 57\,\mu$s. Note this device comes from the 8-inch process, demonstrating compatibility with current semiconductor industry flow. We also fabricated qubit devices (Q2) using the 4-inch process, which yields a $T_1 = 40\,\mu$s, indicating the WJ process can be used in a laboratory cleanroom. 

Figure$\,$5(b) shows the envelope for Ramsey $T_2^*$ and Spin-echo $T_2$ measurements. The measured $T_2$ echo is around $\SI{63}{\us}$. For the frequency tunable qubit Q2, we conducted time-dependent $T_1$ swap spectroscopy (TSS) to extract qubit energy relaxation time as a function of both the qubit operating frequency and time. Such a scan provides information about the spectral density of strongly coupled TLS and about their fluctuation in time. This information may help us improve qubit quality in the future. (See SI \ref{sec: TSS} for more details).

To understand the low $T_1$ results from Sample Q0 (PECVD), we compute the influence of residual SiO$_2$ that remains under the Al counterelectrode of the junction bridge following incomplete removal of the sacrificial oxide, as seen in Fig.$\,$4(d). We treat the lossy SiO$_2$ as a parallel plate capacitor in series with the lossless capacitance from vacuum. The contribution to qubit loss from the residual SiO$_2$ is given by
\begin{align}
Q^{-1}=\frac{C_{pp}}{C_t} \frac{d_{ox}}{d_{pp}}\frac{\delta_{\rm SiO_2} }{\epsilon_{\rm SiO_2}} \ \ .
\end{align}
Here $C_{pp} = \epsilon_0\,A_{pp}/d_{pp} \simeq 0.7\,$fF is the parallel-plate capacitance of the window junction crossover to ground, using an effective area $A_{pp} \simeq 16\,\mu\textrm{m}^2$ and a height $d_{pp} = 0.2\,\mu$m of the suspended airbridge.  The transmon capacitance is $C_t = 80\,$fF.  For SiO$_2$, we take the relative permittivity $\epsilon_{\rm SiO_2} = 3.8$, the loss tangent $\delta_{\rm SiO_2} = 0.005$ and an effective thickness of $d_{ox} = 20\,$nm.  These parameters give $Q \sim 1\,$M, or $T_1 \simeq 30\,\mu$s.  We believe the extreme roughness of the SiO$_2$ residue produces shorter coherence times.  Clearly, clean removal of the SiO$_2$ is a key process factor.

\begin{table}[h!]
\centering
\caption{Transmon qubit parameters.}
\label{tab:QubitProperties}
\begin{tabular}{lccc}
\toprule
\textbf{Qubit Property} & \textbf{Q0} & \textbf{Q1} & \textbf{Q2} \\
%Process(temp. info) & ITRI & ITRI & ITRI+Qolab \\
Type of SiOx & PECVD & PVD & PVD \\
Dry etch SiOx removal & Yes & No & No \\
Qubit frequency & Tunable & Fixed & Tunable \\
Qubit type & Circmon & Floating & Circmon \\
Capacitor gap [$\mu$m] & 20 & 5 & 5 \\
Qubit best relaxation ($T_1$) [$\mu$s] & 2 & 57 & 40 \\
Qubit frequency ($f_q$) [GHz] & 5.23 & 4.34 & 4.57 \\
Dispersive shift ($\chi$) [MHz] & 1.67 & 0.25 & 1.92 \\
Anharmonicity ($\alpha$) [MHz] & 220 & 310 & 200  \\
% \bottomrule
\end{tabular}
\end{table}
%RM: I've commented out the line about Process type. Note that all processing done in the US was done by UW employees, not Qolab employees. No need to explicitly mention this in the paper though.
%Junyi: not sure the highest T1 measurement is 30 or 42us. TBC.

\begin{figure}[htbp]
\includegraphics{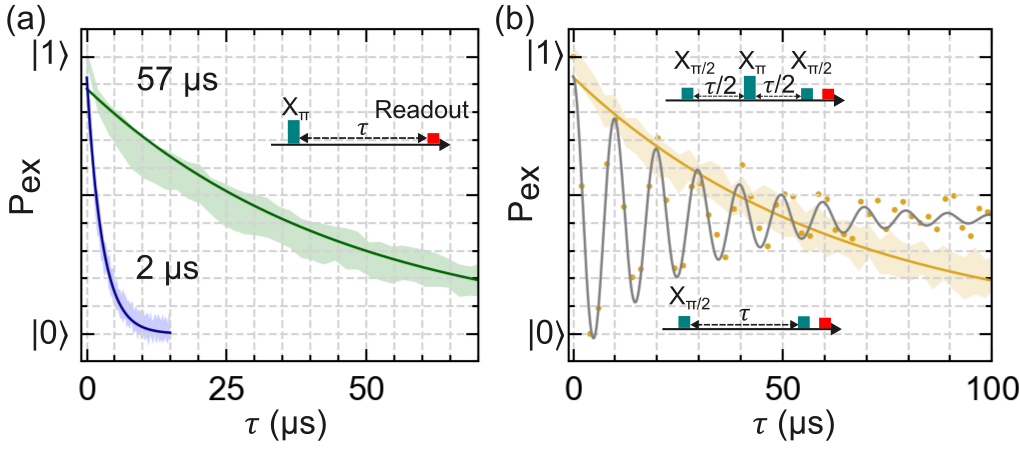}%
\caption{Characterizing qubit performance. (a) Relaxation time $T_1$ measurement for qubit chips without (blue, Q0) and with (green, Q1) the optimal SiO$_2$ process. The curve is the best fitting associated with the deviation in shadow, where we repeated the measurement 50 times and each data point is averaged 500 times. (b) Measurement of the Ramsey dephasing $T_2$*$= 21.2\,\mu$s and spin-echo $T_2 = 63.3\,\mu$s for qubit chip Q1. The Ramsey dephasing only shows the oscillation envelope. }
\end{figure}

So far, we have demonstrated fabrication using window junctions that has the potential to integrate with industry-level processes. The experiments were performed on a limited number of lots, and we have not yet drawn conclusions on qubit yield or the maximum coherence that this fabrication process can achieve. We aim to improve the fabrication yield by optimizing the oxidation condition in the 8-inch system. Additionally, qubit characterization remains crucial; by utilizing time-dependent T$_1$ SWAP spectroscopy, we can investigate coherence loss mechanisms. Clearly, the parallel-plate capacitance can be reduced by shortening the overlap. Lastly, reducing the SiO$_2$ residues through optimzation may be important.  We are planning to perform more experiments to understand yield, maximum coherence, center-to-edge variation, and lot-to-lot variations. 

\section{Conclusion}
Using a removable scaffold, we demonstrated a lift-off-free WJ fabrication process for superconducting qubits. This approach addresses critical challenges in conventional tilted angle deposition for Josephson junction fabrication by mitigating contamination and reducing surface defects. The SiO$_2$ scaffold protects the surface throughout fabrication and can be cleanly removed using vHF process, resulting in minimal surface damage and eliminating the potential issues due to wafer-scale cleaning and oxidation. The results show Josephson junctions with low resistance variation and robust superconducting properties. Resonators with the PVD scaffold process reach an internal quality factor $Q_i$ of $\SI{1}{M}$ for a $12\,\mu$m gap, and qubits devices achieve coherence times $T_1$ up to $\SI{57}{\us}$ for a $5\,\mu$m gap. In contrast, WJ samples using PECVD scaffolds without being fully removed reveal significant degradation in relaxation times, which can be understood using a simple parallel plane capacitor model.

This WJ process integrates seamlessly with industrial fabrication technologies and offers scalability for large-scale qubit manufacturing. These advancements contribute to quantum process science, bridging research innovation and large-scale implementation. By further optimizing the process parameters to address the potential losses associated with residual dielectric materials, the WJ method may pave the way for fault-tolerant quantum computing.

\begin{acknowledgments}
The authors gratefully thank Eric Kao and Britton L. T. Plourde for the fruitful discussion. The authors of Academia Sinica would like to acknowledge funding support from the Academia Sinica Grand Challenge project with project code AS-GCP-112-M01, National Quantum Initiative with project code AS-KPQ-111-TQRB, and NSTC with project code 113-2119-M-001-008. The authors acknowledge Applied Materials for the help on the PVD deposition and ITRI for fabrication support. The authors also acknowledge the fabrication facilities and instrumentation at the Institute of Physics and UW-Madison Wisconsin Centers for Nanoscale Technology.
\end{acknowledgments}

\section{Author Contributions}
R.M. and J.M. designed and simulated the structure of the device. C.T.K., J.Y.T., Y.C.C., P.S.C., C.M.L., A.H., R.M., J.M., and C.D.C. participated in the device's fabrication discussion. E.B., M.A.S., S.W., P.S.C., C.M.L., S.S.S., and Z.Y. conducted the device fabrication. Z.W.X., E.B., M.A.S., and S.W. conducted sample characterizations. C.T.K. and J.Y.T. wrote the first draft of the manuscript, and A.H., R.M., J.M., and C.D.C. provided input and revision. J.Y.T and Y.C.C. coordinated the process integration and data analysis. C.T.K., S.S.S., R.M., J.M., and C.D.C. supervised the project. All authors contributed to the discussion and interpretation of the results.

\bibliographystyle{naturemag}
\bibliography{bibWJ}
\clearpage 

%appendix methods
\section{Supplementary Information}
\beginsupplement

\subsection{Via Fabrication Uniformity}
\label{sec:ViaFabricationUniformity}
To check the uniformity of the via size, we leverage image analysis using Python code to extract the diameter of the via. Two chips with 24 junctions are studied with four designed via sizes, including $\SI{275}{}$, $\SI{300}{}$, $\SI{325}{}$, and $\SI{350}{n\meter}$ in diameter. Fig.$\,$S1 (a), we first take the SEM top view image of the via, finding the edge with the standard edge finding analysis. Next, we fit the edge of the via image with an ellipse function and average the long and short axes as the diameter of the via. The failed fitting outliers are manually excluded due to particle contamination or low image contrast. The ECDF in Fig.$\,$S1 (b) shows the cumulative diameter of 161 successfully analyzed devices with variation under \SI{3.2}{\%}. This process demonstrated a way to make a well-defined via size for the junction region.

\begin{figure}[htbp]
\includegraphics{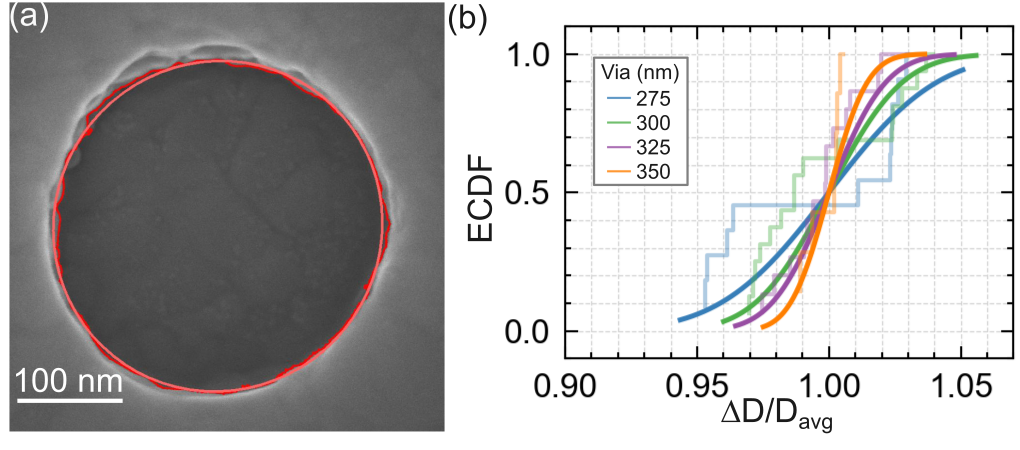}%
\caption{Verification of the via size uniformity. (a) Typical SEM image of a via with diameter \SI{300}{nm}, measuured from a Python image process. The circle is the best ellipse fitting from the image edge detection data, from which we calculate the averaged diameter of the via. (b) The ECDF of the via with the diameter $D$ of \SI{275}{}, \SI{300}{}, \SI{325} and \SI{350}{\nm}. The corresponding weighted average of the via size non-uniformity for each size is \SI{3.2}{\%}, \SI{2.1}{\%}, \SI{1.6}{\%}, \SI{1.1}{\%}, respectively.}
\end{figure}

\begin{figure}[htbp]
\includegraphics[width=\columnwidth]{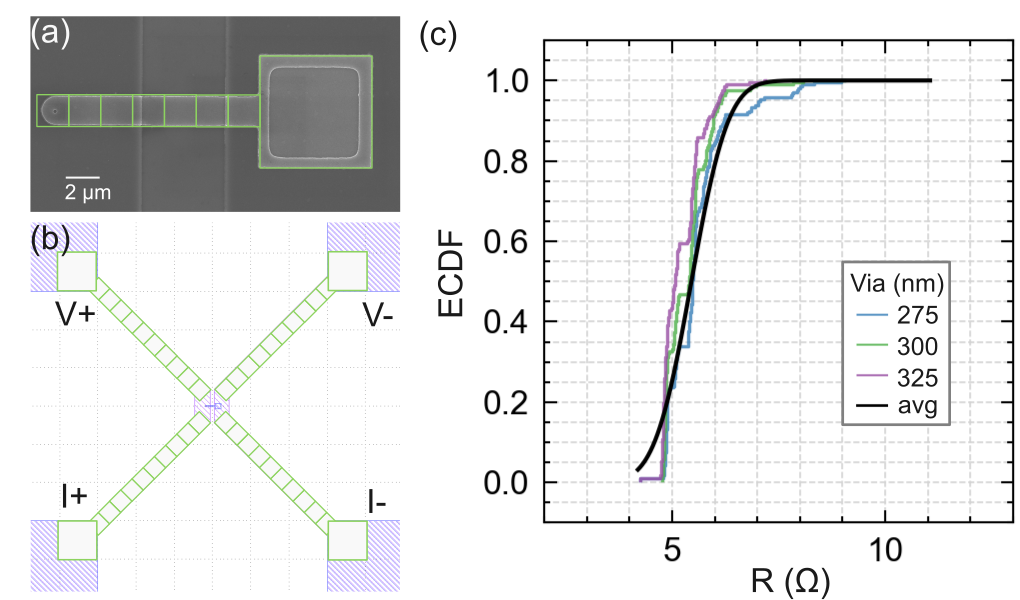}%
\caption{Sheet resistance calculation from the Al film square resistance. (a) A SEM image of the window junction allows us to estimate 9 squares from the top Al electrode. (b) Schematic plot of the 4 probe measurements, where we can calculate a single arm for the resistance consists 11 squares. (c) The ECDF plot for different via sizes of junctions. The contact quality for \SI{300}{} and \SI{325}{n\meter} is similar and \SI{275}{n\meter} shows much higher resistance values. } 
\end{figure}

\subsection{Aluminum Sheet Resistance Calculation and Metallic Contact Testing}
\label{sec:Sup_AlResistance}
To optimize and debug fabrication failures in the window junction process, we designed a test junction layout for the four probe measurement, as shown in Fig.$\,$S2(a). For the test junction structure, a window junction can be fabricated at the center of the base electrode, as depicted in Fig.$\,$S2(b) allowing us to conduct the four-probe measurement. The four arms are started from the upper-left corner and count clockwise as V+, V-, I-, and I+, respectively. Here, we process a wafer with the test junction structure using the window junction fabrication flow and skip the oxidation and vHF process as mentioned in the main context [Fig.$\,$1 (a)]. In Fig.$\,$S2 (c), the ECDF shows the 300 Josephson junctions measured at room temperature with the average resistance \SI{5.7}{\ohm} and the standard deviation \SI{0.4}{\ohm}  (approximately \SI{7}{\%} variation). Theoretically, we can calculate a \SI{100}{n\meter} Al with the resistivity \SI{2.7e-8}{\ohm\meter}, where the resistance per square can be simply calculated from the formula $R_s=\frac{\rho}{t}\approx0.27\Omega/\square$, where $\rho$ is the resistivity of Al and t is the thickness of Al, \SI{100}{n\meter} From Fig.$\,$S2 (a) and (b), we count the number of squares in the geometry: the bridge consists of approximately 9 squares and the total length covers approximately 11 squares, resulting in a total of 20 squares. Therefore, the resistance calculated from this geometry design is \SI{5.6}{\ohm}. Notice that we ignore the contact area of the pillar for the calculation.

\subsection{Transmission Electron Microscopy cross-section study on Window junction}
\label{sec:WJTEM-EDX}
To examine the oxidation process and identify residual elements such as fluorine and argon during fabrication, we conducted a semi-quantitative Transmission Electron Microscopy with Energy-Dispersive X-ray Spectroscopy (TEM-EDS) analysis. Fig.$\,$S3 (a) and (b) show cross-sectional images for the top row without oxidation and the bottom row with oxidation. The atomic weight percentage line scans across the AlO$_x$ barrier were captured with and without oxidation, as shown in the rightmost figures of Fig.$\,$S3 (a) and (b). The line scan and element mapping for the non-oxidized window junction in Fig.$\,$S3 (a) reveals no silicon (Si) or oxygen (O) contamination at the Al interface. Additionally, the cross-sectional TEM confirms that the mild Ar ion milling process effectively removes native AlO$_x$ and post-etching residues without causing damage to the aluminum surface. For the oxidized window junction, the cross-sectional TEM image shows a flat Al-AlO$_x$-Al structure with a thin ($\approx$ \SI{1}{n\meter}) aluminum oxide layer sandwiched between two Al layers. Here we note that to create a clear contrast the WJ for the TEM sample is fabricated on SiO$_2$/Si substrate, hence, there are Si and O signals below the base Al electrode.  

\begin{figure}[htbp]
\includegraphics{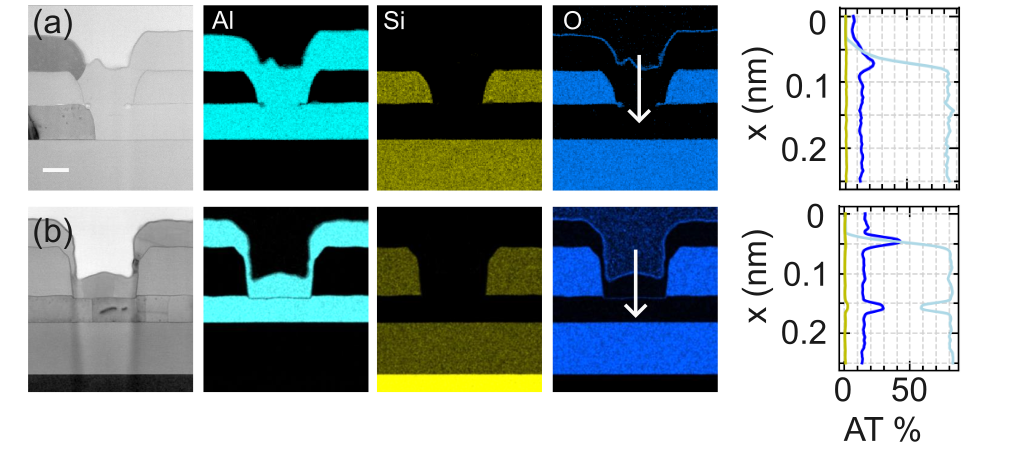}%
% data from ITRI #275_250 
\caption{TEM-EDS analysis corresponding to Fig.$\,$2(c). The TEM images are paired with element density maps (Al, Si, and O) and atomic weight percentage line scans for the non-oxidized (a) and oxidized (b) window junctions. In both cases, fluorine and argon signals are at noise levels and are therefore not shown. The arrow on the oxygen (O) map indicates the direction of the atomic weight (AT\%) line scan across the AlO$_x$ barrier. The line scan confirms the successful formation of the aluminum oxide layer, even within the small via region. The scale bar in (a) is \SI{100}{n\meter} and applies to all images in Fig.$\,$S3.}
\end{figure}

\subsection{Microwave Measurement Setup}
This section presents the microwave setup of our cryogenic system. Figure$\,$S4 illustrates the measurement wiring diagram. All samples were measured using a similar amount of attenuation. In the read-out chain, we have two stages of amplification. An LNF HEMT amplifier is on the \SI{4}{\kelvin} plate with an amplification of \SI{44}{\dB}. Another stage of amplification is at room temperature, an LNF amplifier with an amplification of \SI{40}{\dB}. The room temperature instrument is not included in the diagram. 

\begin{figure}[htbp]
\includegraphics{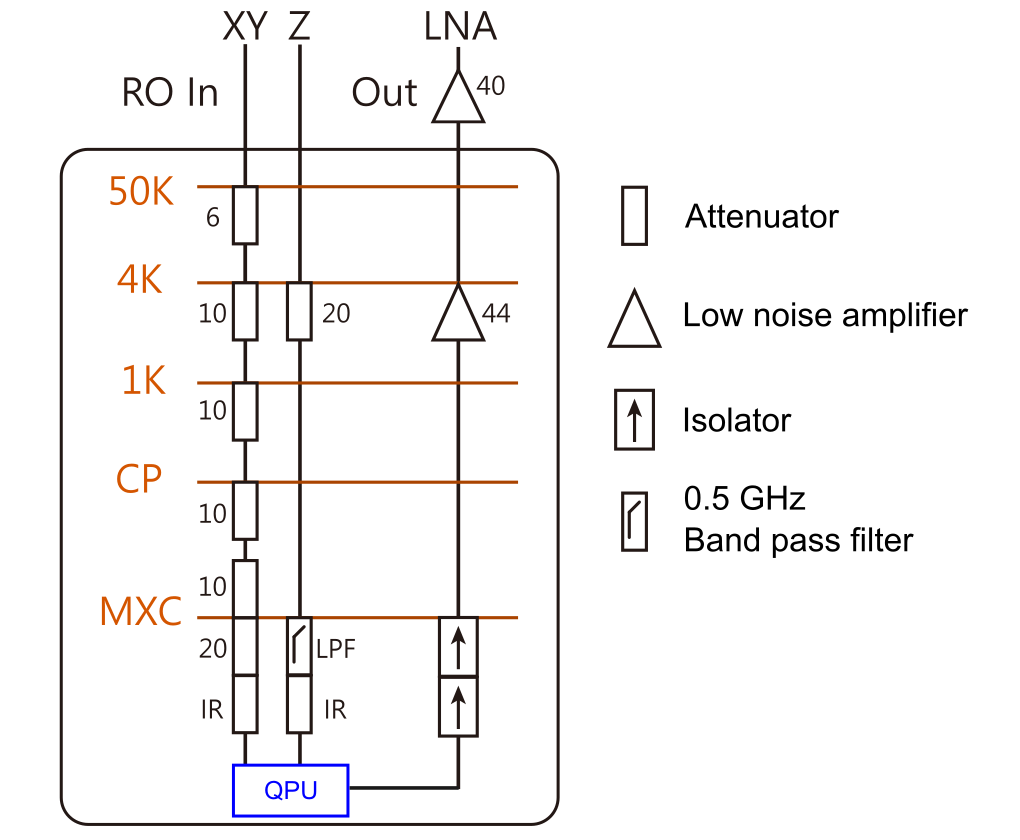}%
\label{sec: MicrowaveMeasurementSetup}
\caption{We sketch the low-temperature microwave setup for the qubit chip characterization. The room temperature RF electrics are not presented. The labeled numbers next to the component indicate the amount of attenuation in dB.}
\end{figure}

\subsection{Aluminum Surface Analysis}
This section shows studies of the Al surface before and after vHF treatment. The top row in Fig.$\,$S5(a) and (b) are SEM and AFM images on the Al surface, as deposited. One can clearly observe the grain of Al. In Fig.$\,$S5(b), the AFM image reveals an RMS roughness of \SI{1.2}{n\meter}. Here we note that all four images share the same scale bar e.g. the same area size of \SI{1}{\mu\meter} by \SI{1}{\mu\meter}. After vHF treatment to remove the SiO$_2$ scaffold, we verify the Al surface as presented in Fig.$\,$5S(c) and (d) for SEM and AFM images, respectively. The Al surface remains intact and shows the Al grain without damage under the SEM image. From Fig.$\,$5S(d), we extracted a similar surface RMS roughness of \SI{1.5}{n\meter} indicating that the surface roughness is comparable with pristine Al. However, the reduced grain size in AFM image suggested that there may be a thin layer of SiO$_2$ remaining resulting in a smaller grain feature. This thin layer of SiO$_2$ may be transparent in SEM and therefore could not be seen properly using an electron beam. This suggests that we may need to improve the vHF process and could potentially yield a better T$_1$ in the future.  

\begin{figure}[htbp]
\includegraphics{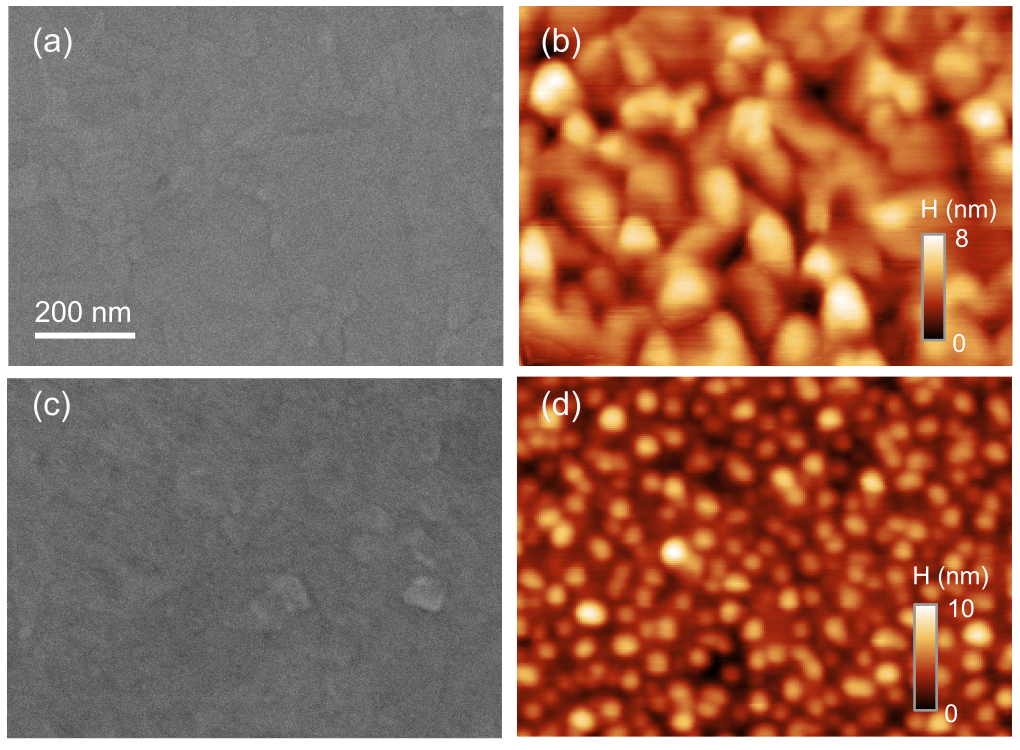}%
\label{sec: AlRoughness}
\caption{Measurement of the Al surface roughness from a blanket Al film grown on the silicon substrate. (a) SEM of the Al surface as deposited, and (b) corresponding AFM data. (c) and (d) SEM and AFM data of the Al surface after SiO$_2$ deposition and vHF removal.}
\end{figure}

\subsection{SWAP spectroscopy}
Here, we conduct a time-dependent $T_1$ spectroscopy (TSS) as a function of the qubit frequency. TSS shows the $T_1$ behavior of the WJ qubit to the surrounding energy relaxation\cite{klimov_fluctuations_2018, niu_learning_2019}. The strong $T_1$ relaxation regions could be ascribed to TLS or telegraphic modes that interact with qubits as shown in Fig.$\,$6S(c). A sliced TSS at t=\SI{0}{} and \SI{7}{\hour} plotted as a function of the relaxation time $T_1$ identifies several of the local minimum $T_1$ labeled with a black arrow. Previous results show such behavior could correlate to the qubit-TLS coupling.\cite{klimov_fluctuations_2018}; therefore, the $T_1$ is shorter than \SI{1}{\us} providing information of the TLS. Using the TSS can help us identify the possible TLSs during the WJ process aiming further to improve the $T_1$ of the WJ qubits. 

\begin{figure}[htbp]
\includegraphics{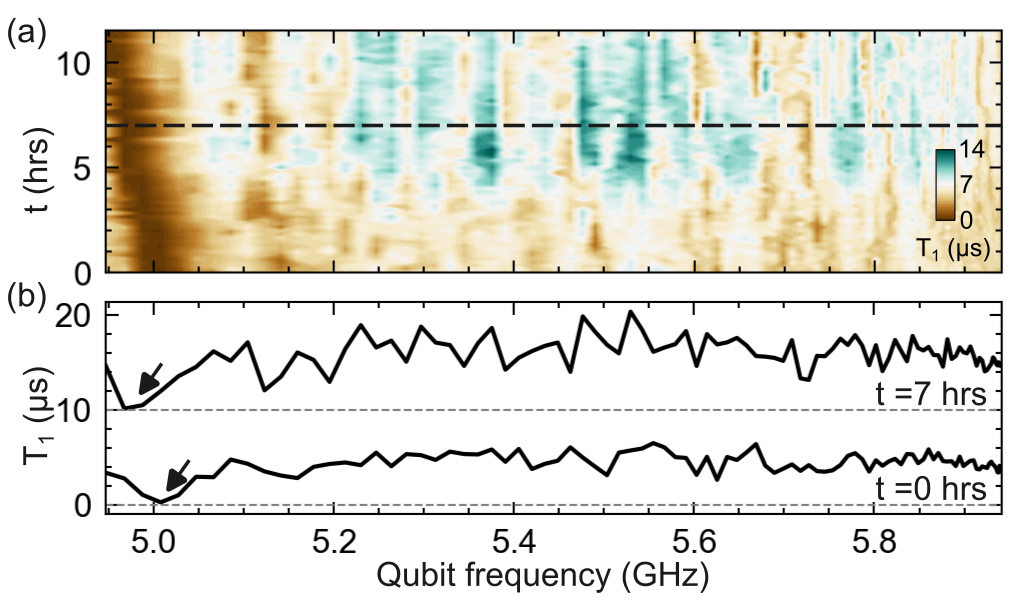}%
\label{sec: TSS}
\caption{Time-dependent swap spectroscopy(TSS) of Q2 (Table I.). (a) The time-dependent swap spectroscopy (TSS). (b) Sliced T1 spectroscopy at t=0 hours and t=7 hours from (a).}
\end{figure}

%% Below are all for backup%% Below are all for backup%% Below are all for backup
%% Below are all for backup%% Below are all for backup%% Below are all for backup
%% Below are all for backup%% Below are all for backup%% Below are all for backup
%% Below are all for backup%% Below are all for backup%% Below are all for backup
% \clearpage %new page
% \FloatBarrier %new page

% \subsection{Aluminum Surface Roughness}
% Remove this section. This is for the Argon plasma cleaning process, not for the ion-milling process check. I just put here for the ref. 
% \label{sec: AlRoughness}
% \begin{figure}[htbp]
% \includegraphics{Fig/Sup_AlAFM.png}%
% \caption{TBD} 
% \end{figure}

% \subsection{qubit coherence measurement}
% The schematic in the main text Figure 4(a) shows the pulse sequence for $T_1$ measurement. We prepare the qubit in the ground state and excite the qubit with the $\pi$ pause from the feedline, waiting with time $\tau$ for the readout measurement from the feedline. Each point is averaged 500 times, and then the $T_1$ time is extracted with the $T_1$ fitting. Here, we repeated 100 times to extract the $T_1$ measurement. 

% Similar to the $T_1$ measurement, the Ramsey measurement is that we excite the qubit with $\pi/2$ and wait for time $\tau$. The spin-echo method is measured with the sequence pulse including a $\pi$ pulse between two $\pi/2$ pulses with $\tau/2$ delay time.

\end{document}